\PassOptionsToPackage{pdftex,pdfencoding=auto,colorlinks=true,urlcolor=blue,linkcolor=blue,citecolor=blue}{hyperref}
\documentclass[floatfix,%
 reprint,
superscriptaddress,
 amsmath,amssymb,
 aps,
pra,
]{revtex4-2}
\usepackage[UKenglish]{babel}
\usepackage{latexsym}

\usepackage{graphicx}	
\usepackage{amssymb}
\usepackage{amsmath}
\usepackage{xcolor}
\usepackage[normalem]{ulem}

\pdfoutput=1

\usepackage{bm}
\usepackage{verbatim}
\usepackage{amsmath}
\usepackage{amssymb}
\usepackage[utf8]{inputenc}
\usepackage{nicefrac}
\usepackage{braket}
\usepackage{pdfcomment}

\usepackage{pbox}
\usepackage{makecell}
\usepackage{hhline}
\usepackage{multirow}
\usepackage{tabularx}
\usepackage{soul}
\DeclareUnicodeCharacter{0308}{HERE!HERE!}
\usepackage{graphicx}
\usepackage{indentfirst,csquotes}

\usepackage{amssymb,amsthm,amsmath}
\usepackage{xcolor,paralist,titlesec,fancyhdr,etoolbox}

\usepackage{physics}

\titleformat{\section}[display]{\normalfont\huge\bfseries\centering}{\centering\chaptertitlename\thechapter}{10pt}{\Large}
\titlespacing*{\section}{0pt}{0ex}{0ex}

\hypersetup{ colorlinks=true, linkcolor=black, filecolor=black, urlcolor=black }

\usepackage{lipsum}
\usepackage[colorlinks=true,urlcolor=blue,linkcolor=blue,citecolor=blue]{hyperref}
\begin{document}
%############################## TITLE #########################################
\title{Magnetic brightening of light-like excitons in a monolayer semiconductor}
%##############################################################################
%
%############################ AUTHORS #########################################

%
\author{A. Delhomme}
\affiliation{Walter Schottky Institut and TUM School of Natural Sciences, Technische Universit\"at M\"unchen, Am Coulombwall 4, 85748 Garching, Germany.}
\author{T. Amit}
\affiliation{Department of Molecular Chemistry and Materials Science, Weizmann Institute of Science, Rehovot, Israel.}
\author{P. Ji}
\affiliation{Walter Schottky Institut and TUM School of Natural Sciences, Technische Universit\"at M\"unchen, Am Coulombwall 4, 85748 Garching, Germany.}
\author{C. Faugeras}
\affiliation{Laboratoire National des Champs Magnetiques Intenses, CNRS-UGA-UPS-INSA-EMFL, 38042 Grenoble, France.}
\author{S. Refaely-Abramson}
\affiliation{Department of Molecular Chemistry and Materials Science, Weizmann Institute of Science, Rehovot, Israel.}
\author{J. J. Finley}
\affiliation{Walter Schottky Institut and TUM School of Natural Sciences, Technische Universit\"at M\"unchen, Am Coulombwall 4, 85748 Garching, Germany.}
\author{A. V. Stier}
\affiliation{Walter Schottky Institut and TUM School of Natural Sciences, Technische Universit\"at M\"unchen, Am Coulombwall 4, 85748 Garching, Germany.}
%
%##############################################################################
%
\date{\today}
%
%##############################################################################
%									ABSTRACT
%##############################################################################
%
\begin{abstract}
Monolayer transition-metal dichalcogenides, such as WSe$_2$, are direct gap, multi-valley semiconductors. Long-range electron-hole exchange interactions mix the valleys, yielding dispersion relations for massive ($\propto Q^2$) as well as light-like ($\propto Q$) excitons.
We report magneto-photoluminescence spectroscopy of excitons in the monolayer semiconductor WSe$_2$ to $B = \pm25$T. The magnetic field-dependent line shape of the neutral exciton reveals the emergence of a new blue-detuned emission peak in both field orientations. Analyzing the distinct magnetic field-dependent shifts of both peaks facilitates the identification of the emergent feature as a spin-singlet with a significantly smaller reduced exciton mass as compared to the neutral exciton. The intensity of the emergent feature increases with magnetic field according to $\propto B^2$, as expected for a linear dispersion relation. The density-dependent diamagnetic shift ratios of both features follow the expected density dependence of the electron-hole exchange interactions. We interpret our observations within a picture of magnetic-field-induced coupling between the bright massive and quasi dark light-like exciton, leading to its brightening.  
\end{abstract} %%%%%%%%%

\maketitle
\begin{figure}
    \centering
    \includegraphics[width=1.\columnwidth]{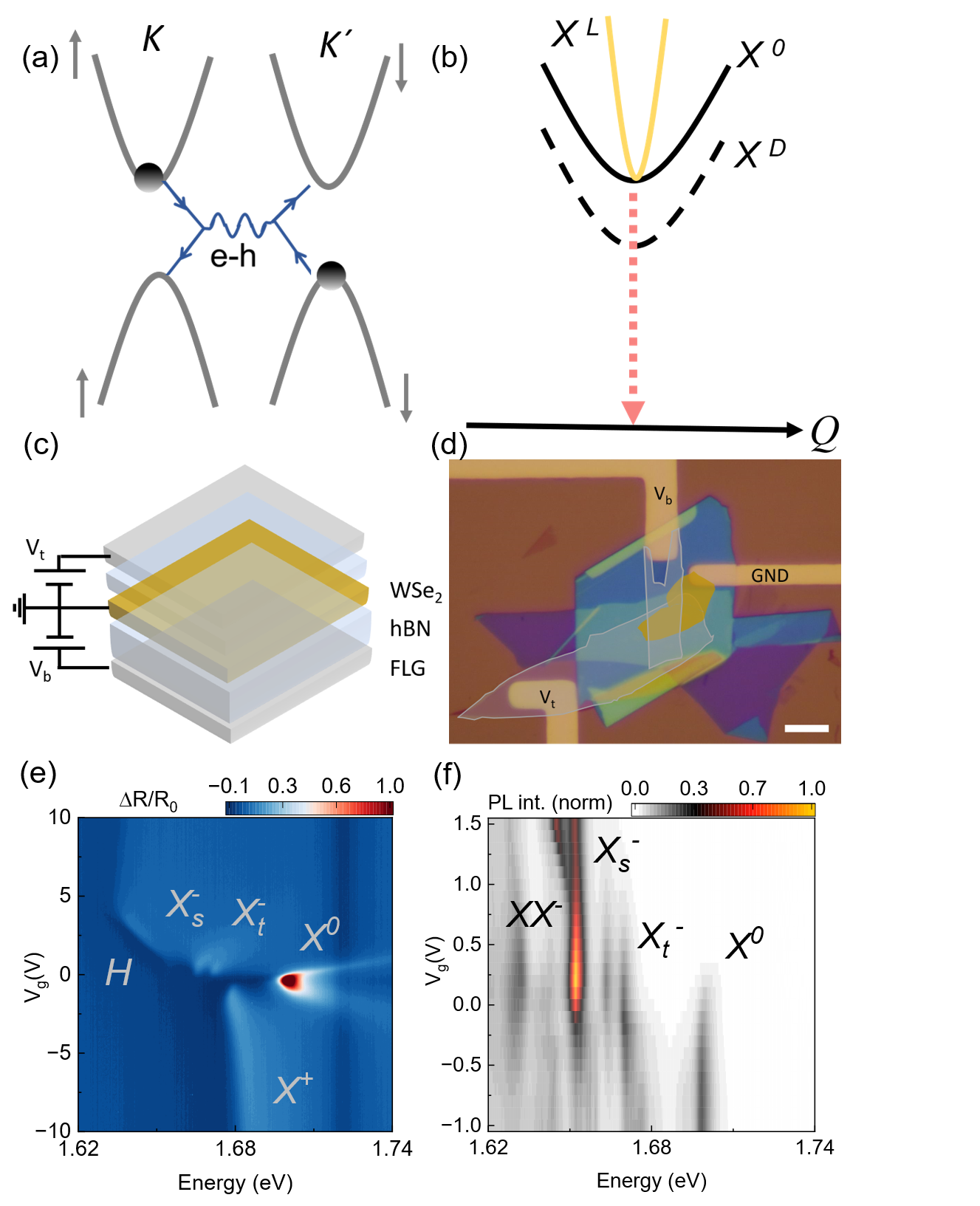}
    \caption{a) Schematic of the K/K' band dispersion of a monolayer TMD in the single particle picture, indicating the valley coupling through the electron-hole exchange interaction. b) Schematic of the exciton dispersion relations of bright ($X^0$), dark ($X^D$) and light-like ($X^L$) excitons. c) Schematic of the device structure of a dual gated monolayer WSe$_{2}$. d) Optical image of the device, indicating top(bottom) gate contacts and the ground. Few layer graphene (FLG) gates are indicated by the light grey shaded regions, the WSe$_{2}$-ML is inducated in yellow. e) Differential reflectance map of the device showing neutral and charged excitons. f) Normalized PL specta of the WSe$_2$ monolayer. The excitation is with a $633~nm$ solid state laser diode, unpolarized and focused on the sample using a high numerical aperture ($NA=0.82$). Dominant PL features are the neutral exciton $X_0$, the negatively charged biexciton $XX^-$, the triplet $X^-_t$ and singlet trions $X^-_s$ and the charged biexciton $XX^-$.}
    \label{figure1}
\end{figure}

%
%###############################################################################
%								MAIN TEXT
%###############################################################################
%
Coulomb-correlated, photo-excited electron-hole pairs in semiconductors, excitons, have been a focus of investigation for almost a century \cite{Frenkel31,Wannier37,Mahan67,haug2009quantum}. In undoped, direct-gap semiconductors, such states form from an electron in the conduction band (CB) and a hole in the valence band (VB) and usually manifest as a discrete optical resonance below the single particle band gap \cite{Elliott57}. Coulomb interactions in two dimensional (2D) transition-metal dichalcogenides (TMDs) are significantly enhanced as compared to 3D semiconductor counterparts due to the extreme quantum confinement and reduced dielectric screening. The Coulomb interactions $K^{eh}=K^d+K^x$ consist of a direct term ($K^d$) describing the screened attractive interaction between the electron and the hole and an exchange term ($K^x$), where the exchange scattering of an electron-hole pair gives rise to repulsive interactions \cite{Qiu.2021}.  
The very strong light-matter coupling in monolayer semiconducting TMDs such as MoS$_2$ and WSe$_2$ originates in part from their direct band gap and large exciton binding energies (E$_B$) \cite{Mak.2010,Splendiani2010,Wangcolloquium2018} due to $K^d$.  As a consequence of steady improvements in material quality and sample design, a plethora of excitonic resonances have been investigated and identified across many 2D semiconductors. These include the Rydberg states of the neutral exciton \cite{Chernikov.2014,Stier.2018}, multi-exciton complexes such as biexcitons \cite{Barbone.2018} and even higher multi-particle excitations \cite{He2020,VanThuan2022} as well as various charged excitons in electron- or hole doped samples \cite{Heinz2012b,Ross2013,Jones.2015,Plechinger.2016,Grzeszczyk.2021,Klein2022,Leisgang2024}. Adding to the richness of excitonic species in this material system is the multivalley band structure, depicted in Fig.~\ref{figure1}a, where due to broken inversion symmetry and strong spin-orbit coupling, time reversal pairs of direct bandgaps at the $K/K'$-points of the Brillouin zone form in the monolayer limit \cite{Xiao.2012, Splendiani2010, Mak.2010}. This leads to the situation where in addition to the bright excitons, spin- and momentum forbidden dark excitons can form from electron-hole pairs characterized by permutations of the spin- and valley quantum numbers of the constituent particles \cite{Molas.2017,Zhang.2017,Malic.2018,Robert.2020,Qian2024}. 
When the electron-hole exchange interactions $K^x$ are taken into account, excitons are never pure valley eigenstates but have quantum states that are superpositions of $K$ and $K'$ \cite{Qiu.2015, Qiu.2021, Klein2022}. For excitons having finite center of mass momentum ($Q\neq0$), it is theoretically predicted that the long range exchange interactions give rise to an additional exciton dispersion \cite{Qiu.2015}, the light-like exciton $X^L$, schematically depicted in Fig. \ref{figure1}b. $X^L$ is energetically degenerate with the massive exciton ($X^0$, $E(Q)\propto Q^2/2M$,$M=m_e+m_h$) at $Q=0$. Like the massive exciton, it is a spin-like optical transition \cite{Qiu.2021}. However unlike $X^0$, its dispersion relation gains a linear term in 2D ($X^L$, $E(Q)\propto Q^2/2M+\alpha Qcos^2(\theta_{Q})$), where $\alpha=\frac{\hbar^2}{m^2}\lvert\sum_{vc\textbf{k}}(E_{c\textbf{k}}-E_{v\textbf{k}})^{-1}A^s_{vc\textbf{k}}\bra{c\textbf{k}}\textbf{p}\ket{v\textbf{k}}\lvert^2$ describes all band-to-band transitions comprising the exciton and $\theta_Q$ is the angle between the momentum $Q$ and the exciton transition matrix element.\cite{Qiu.2021} When the exchange interaction is strong, the linear-in-Q term dominates and therefore, the $X^L$ dispersion is light-like. This state has recently been experimentally observed in hexagonal boron nitride using momentum-resolved electron energy-loss spectroscopy.\cite{liu.2025}  In optical spectroscopy, the conventional massive neutral exciton branch has been used to explain all experimental observations. Naturally, only excitons in the light-cone close to $Q=0$ are optically active. The light-like exciton branch features a linear-in-$Q$ dispersion relation and therefore, the joint density of states vanishes close to $Q=0$ and, thus, the exciton is effectively dark. However, the situation changes, when the excitons are not at rest but have finite momentum. Then, massive and light-like excitons mix with a matrix element $\propto Q$ \cite{Lo.2021} and the light-like branch is expected to gain oscillator strength proportional to the $\propto Q^2$ increase of the joint density of states of the linear dispersion relation.\\     
In this Letter, we employ magneto-spectroscopy in high magnetic fields to $B=25T$ to magnetically brighten the light-like exciton branch. 
In general, the identification of all excitonic states in the various 2D materials and van der Waals heterostructures is a matter of ongoing research and heavily relies on investigating the circularly polarized ($\sigma^{\pm}$) optical selection rules and related spin-valley physics. \cite{Xiao.2012} Here, a key tool are high magnetic fields, because the spin-valley degeneracy can be lifted and excitons can be properly identified as each state shifts distinctly with $B$. Importantly, the shifts directly reveal fundamental parameters such as the exciton size, mass and spin, when benchmarked against suitable theoretical models. \cite{Evans67dia,Miura2003,Stier.2016,Stier.2018,Goryca.2019} For example, consider a 2D semiconductor in the weak field limit, where the magnetic length $l_B=\sqrt{\hbar/e B}$ ($\approx 26 nm/\sqrt{B}$) is much larger than the in-plane exciton radius $r_{\perp}$, with expectation value $\expval{r_{\perp}}{\Psi}$ calculated from the exciton envelope wavefunction ${\Psi(r)}$. In this regime, the 1s exciton energy shift in a magnetic field is
\begin{equation}
\Delta E(B)=E_{VZ}(B)+E_{dia}(B)=\pm \frac{1}{2} g_V \mu_B B+\sigma^z B^2, 
\end{equation}
where $\sigma^z =\frac{e^2}{8 \mu}\expval{r_{\perp}^2}$ is the diamagnetic shift coefficient. The valley Zeeman shift ($E_{VZ}$) is due to the valley-dependent magnetic moment, parameterized by the valley g-factor $g_V$ \cite{Yilei,Macneill,Aivazian.2015,Srivastava.2015,Stier.2016,Wozniak.2020,Deilmann.2020}. This term has been particularly important for the identification of various intra- and interlayer excitons as well as distinguishing bright and dark states \cite{Wangcolloquium2018, Robert.2020}. The second term is the quadratic diamagnetic shift, which is proportional to the ratio of the exciton wavefunction threaded by the magnetic field ($\sim \expval{r_{\perp}^2}$) and the reduced exciton mass $\mu=(m_e^{-1}+m_h^{-1})^{-1}$. This term has been particularly important for the unique identification of the exciton Rydberg states, which are directly revealed via their distinct sizes and, therefore, diamagnetic shifts \cite{Evans67dia,Stier.2016, Stier.2018}. Most importantly for this Letter, the diamagnetic shift causes the 1s ground state of the neutral exciton to acquire kinetic energy and consequently finite momentum that scales linearly with magnetic field ($Q\propto B$). As discussed in detail below, we identify the emergence of the light-like exciton via the valley Zeeman and diamagnetic shift in combination with the magnetic field induced brightening.

\begin{figure}
    \centering
    \includegraphics[width=1\columnwidth]{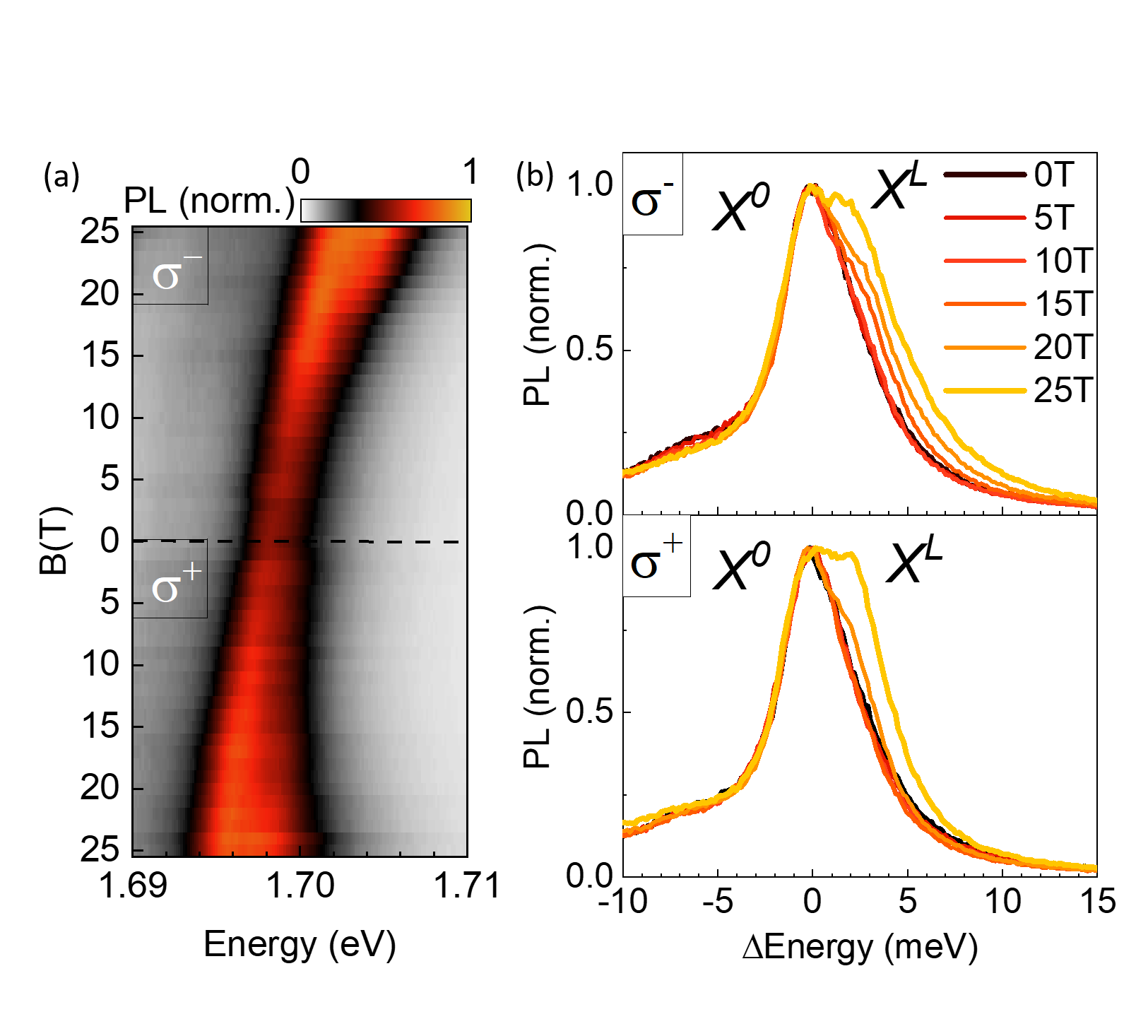}
    \caption{a) False color map of the magneto-PL intensity in the vicinity of $X^0$ in the charge neutral regime at $V_g=-1V$. At high magnetic fields, the PL linewidth apparently broadens towards the blue side.  b) Normalized PL spectra at select magnetic field for $\sigma^\pm$ circular polarization. The energy axis is centered at the $B$-dependent position of $X^0$, highlighting the emergence of $X^{L}$ with increasing magnetic field.}
    \label{figure:2}
\end{figure}

 The investigated van der Waals heterostructure is schematically depicted in Fig.~\ref{figure1}c and an optical image is shown in Fig.~\ref{figure1}d. A monolayer tungsten diselenide (ML-WSe$_2$) flake is fully encapsulated in hexagonal boron nitride (hBN) using dry viscoelastic stamping techniques. The thickness of the top(bottom) hBN layers is $15(30)~nm$, determined by optical contrast and AFM. Top and bottom gate electrodes are fabricated by few-layer graphite (FLG). Electrical contacts to the gates and the ML-WSe$_2$ are defined using e-beam lithography and metal evaporation. The electrical control of the sample is achieved using two voltage sources to independently adjust the potentials of the top ($V_t$) and bottom ($V_b$) gate with respect to drain (GND). In Fig.~\ref{figure1}e, we present typical gate-dependent ($V_g=V_t-V_b$) reflectance contrast of the sample at $T=5~K$, recorded in vacuum with a white light source, and identify the well-known neutral exciton, the positively and negatively charged excitons as well as the hexciton.\cite{VanThuan2022} 
 Magneto-optical spectroscopy was performed in a $20~MW$ resistive DC magnet at $T=4.2~K$ equipped with a helium exchange gas cryostat. %The sample was excited with a $633~nm$ solid state laser diode. The excitation is unpolarized and focused on the sample using a high numerical aperture ($NA=0.82$) objective, which is apochromatic between $500$ and $850~nm$. The sample was positioned with an X-Y-X piezo stage and the excitation beam spot is diffraction limited ($\sim 1\mu m$). The photoluminescence (PL) of the sample is collected via free-beam optics in a $500~mm$ focal length spectrometer equipped with a nitrogen-cooled charge coupled device camera. The circular polarization of the PL signals is analyzed using a combination of a quarter-wave plate and a linear polarizer.
 The photoluminescence (PL) data presented in the false color map Fig.~\ref{figure1}f are recorded at zero applied magnetic field with vanishing net electric field across the ML ($V_g=V_t-2 V_b$). The full width at half maximum of the neutral exciton PL peak is $\sim 4.5~meV$, in line with the finite spatial resolution of high-field magneto-spectroscopy setups. ~\cite{Delhomme2019,Koperski2018,Zhang.2017,Mitioglu2015} Notably, the singlet and triplet trions ($X^-_s$ and $X^-_t$) are well resolved.~\cite{Boulesbaa2015,Plechinger.2016,Molas2017,Jadczak2017,Kapu2020} The map shows electrostatic doping the sample, from the charge neutral to the electron-doped regime. At $V_g=-1V$, the low intensity of the charged exciton features and the dominance of $X^0$ as well as localized exciton features in the PL spectra indicate that the ML is close to charge neutrality. Increasing the applied gate voltage electrostatically dopes the ML-WSe$_2$ with excess electrons, resulting in the progressive quenching of the $X^0$ peak and the concomitant appearance of the charged exciton features $XX^-$, $X^-_s$, and $X^-_t$.~\cite{Barbone.2018} At high $n$- doping above $V_g=+1V$, the PL spectrum shows the emergence of the hexciton-feature (H) arising from the occupation of the second WSe$_2$ conduction band.~\cite{VanThuan2022} 
We now focus on the $B$-field shift of the neutral exciton close to charge neutrality. The false color map in Fig. \ref{figure:2}a shows the circularly analyzed exciton PL as a function of the magnetic field from $B=-25T$ to $B=+25T$ at $V_g=-1V$ in the Faraday configuration (the applied magnetic field $B$ is perpendicular to the sample plane). 

At low magnetic fields $\abs{B}<15T$, the peak of the $\sigma^{\pm}$ polarized exciton emission shifts essentially linearly with $B$, asymmetrically around the zero-field emission energy. This is fully consistent with the expected valley Zeeman effect of $X^0$\cite{Aivazian.2015,Srivastava.2015,MacNeill.2015,Stier.2016,Stier.2018} and we deduce a valley Zeeman g-factor for the $X^0$-peak of $g_V^0=-4.2\pm0.1$. At higher $B$- fields, a superlinear blue shift superimposed on the linear behavior can be observed from the raw data, consistent with expectations from the $\propto B^2$-dependent diamagnetic shift.~\cite{Stier.2016,Stier.2018} Most notably, however, and the key result of this Letter, is the apparent broadening of the emission with increasing $B$. This contrasts strongly with expectations and experimental observations of exciton linewidth \textit{narrowing} with increasing $B$ typically observed in \textit{absorption} experiments.~\cite{Stier.2018}

We investigate the evolution of the PL spectra with increasing positive and negative $B-$field in Fig. \ref{figure:2}b close to charge neutrality. 
The exciton PL intensity is normalized and the energy scale is centered to the position of $X^0$ at the particular magnetic field, i.e. the Zeeman and diamagnetic shift of $X^0$ are subtracted. With increasing magnetic field, an asymmetric shoulder clearly emerges on the high energy side of the exciton, evolving into a distinct peak for the highest magnetic fields above $\sim 20~T$. As we show below, we interpret this new peak as the magnetic field brightened linear dispersive exciton, $X^{L}$. Note that the low energy shoulder of $X^0$ remains unaffected by the magnetic field. This behavior is universal with gate voltage and presented and discussed in Fig. S1 of the Supplemental Material.  
We observe that $X^{L}$ emerges with two notable characteristics: first, it always appears on the high energy side of the neutral exciton and second, $X^{L}$ emerges at lower fields for $\sigma^-$ as compared to $\sigma^+$. The former observation is consistent with the magnetic field dependent splitting between $X^{0}$ and $X^{L}$, $\Delta_{0L}=E_{X^L}-E_{X^0}=\Delta E_{VZ}+\Delta E_{dia}$ being dominated by the difference of the diamagnetic shifts ($\Delta E_{dia}$), while the difference in the valley Zeeman shifts ($\Delta E_{VZ}$) is not the dominant term at high magnetic field.

The asymmetic emergence of $X^{L}$ cannot arise from $\Delta E_{dia}$, but from $\Delta E_{VZ}$, which is anti-symmetric with respect to zero field. This difference in the valley Zeeman shift implies that $X^0$ and $X^{L}$ have different $g-$ factors, which in turn implies that they originate from different excitonic dispersions.~\cite{FariaJunior2019PRB} 
%Therefore, we consider $X^{L}$ as an excitonic peak distinct from $X^0$ giving rise to its own declination of charged excitonic complexes.
\begin{figure}
    \centering
    \includegraphics[width=1.\columnwidth]{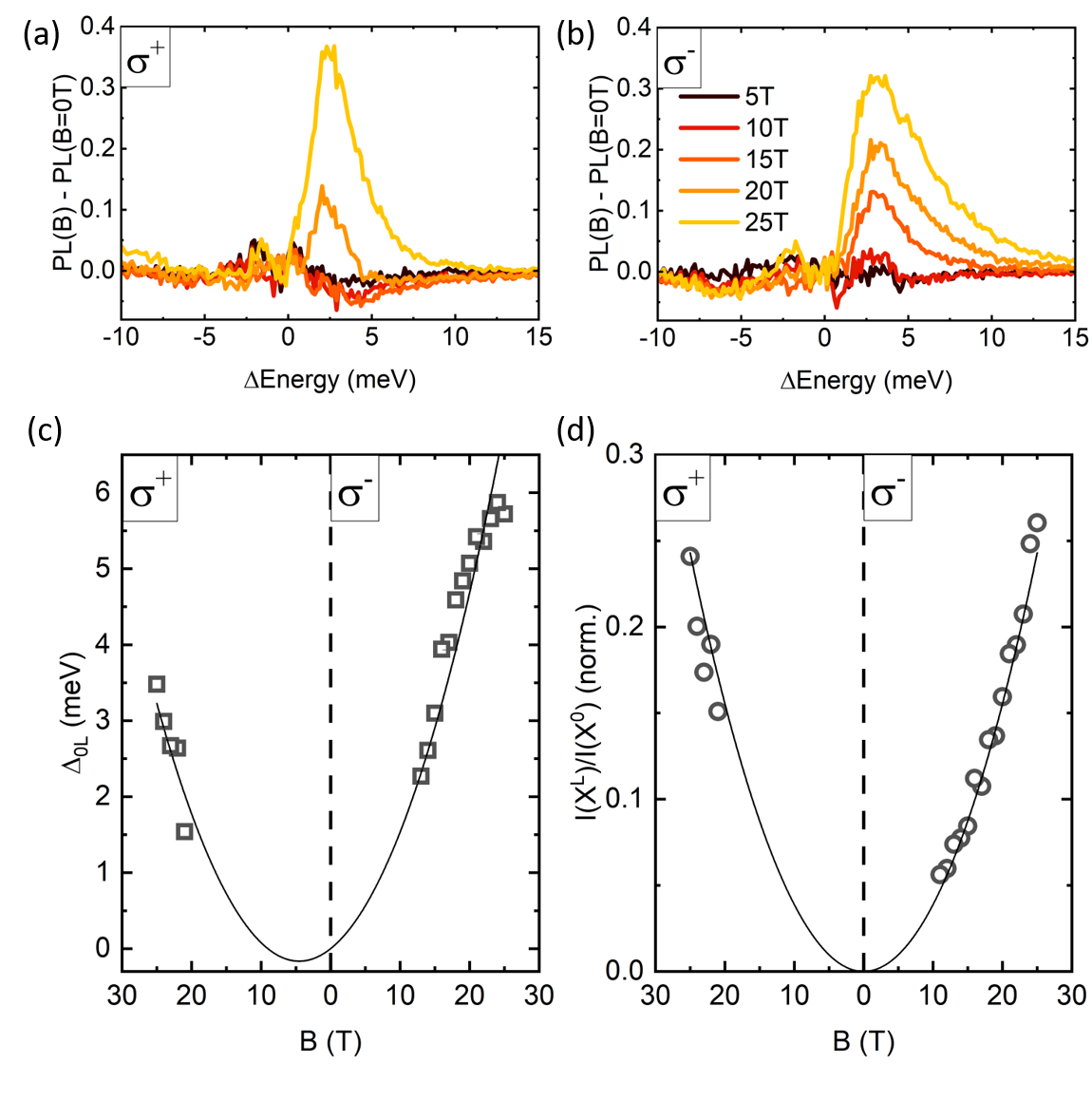}
    \caption{a) Normalized PL spectra at select magnetic fields with the $B=0T$ spectra subtracted as the background for $\sigma^+$ and b) $\sigma^-$ circular polarization. c) Energy separation between $X^0$ and $X^L$ as a function of magnetic field (black squares). Solid line is a 2nd order polynomial fit through the data, revealing the difference in diamagnetic shift and valley Zeeman shift between $X^0$ and $X^L$. d) Ratio of the integrated PL intensity of $X^L$ and $X^0$, revealing $\propto B^2$ brightening of $X^L$.}
    \label{figure:3}
\end{figure}

\begin{figure}
\centering
\includegraphics[width=1.\columnwidth]{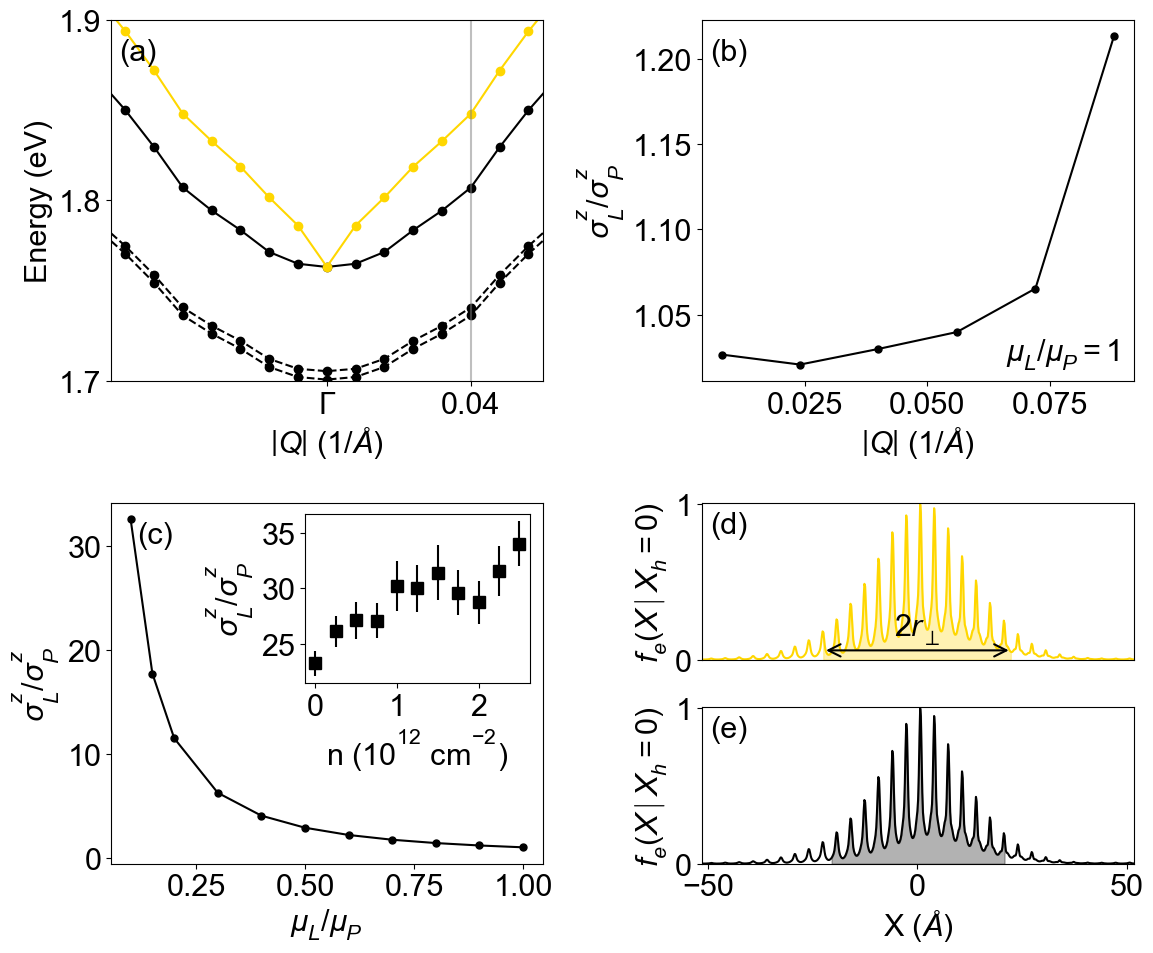}
    \caption{(a) Excitonic bandstructure computed using GW-BSE. Dashed black lines indicate $X^D$; solid black and yellow lines represent $X^0$ and $X^L$, respectively. The grey vertical line marks the exciton momentum $\mathbf{Q}$ analyzed in (c-e). (b) Calculated diamagnetic shift ratio of the two bright excitons for a perpendicular external magnetic field as a function of the exciton momentum $\mathbf{Q}$, using the same effective mass for both states. (c) Calculated diamagnetic shift ratio of the two bright excitons for a perpendicular external magnetic field as a function of the effective mass ratio between the same two excitons. The electron density is computed using a capacitor model with an effective dielectric constant for the hBN of $\epsilon_{hBN}=3.5$. (d) Electron density along the X-axis, integrated over all other directions, for $X^L$. The shaded area shows where 95\% of the electron density is localized, giving an estimated diameter of $44.9\AA$. (e) Same as (d) but for $X^0$, giving an estimated diameter of $41.4\AA$.}
    \label{fig:theory}
\end{figure}

We now continue to quantify these observations by analyzing the magnetic field dependent evolution of the lineshape. The results of this evaluation are presented in Fig. \ref{figure:3}. We use the zero field PL spectrum as a background and subtract the magnetic field dependent PL from it, therefore highlighting the magnetic field induced evolution of the lineshape, and the emergence of $X^{L}$. The background subtracted data are presented in Fig. \ref{figure:3}a and b for the $\sigma^{\pm}$-PL, respectively. To determine the magnetic field-dependent splitting between $X^0$ and $X^L$, we calculate the centroid of the background subtracted PL ($\Delta_{0L}=\sum_{i} \frac{E_iI_i}{I_i}$) by numerical integration across the feature. This is plotted in Fig. \ref{figure:3}c, and can be very well fitted with a second-order polynomial, yielding values for $\Delta g_V=1.3\pm 0.1$ and $\Delta \sigma^z = 6.7\pm0.2~\mu eV/T^2$. Utilizing the valley Zeeman g-factor for the $X^0$-peak ($g_V^0=-4.2$), we deduce $g_V^L=-5.5\pm0.2$. The observed valley Zeeman behavior is universal and independent of gate voltage (see Fig. S2 in the SM). While $g_V^L$ is measurably larger than the $g$-factor of the neutral exciton, $g_{V}^L$ is significantly smaller than the $g$-factor of a dark exciton spin triplet state ($g=-8$).~\cite{Wozniak.2020} As such, we identify $X^0$ and $X^{L}$ to be both spin singlet states (bright states) that arise from different exciton dispersions.\\ 
The diamagnetic shift of $X^L$ is surprisingly large. We determine a diamagnetic shift coefficient for the neutral exciton $\sigma^z_0=0.3\pm0.1\mu eV/T^2$, in line with typical values for diamagnetic shift coefficients of hBN encapsulated ML TMDs of $\sigma^z<0.5~meV/T^2$. \cite{Stier.2018, Goryca.2019,Delhomme.2020} As such, the diamagnetic shift coefficient of the linear dispersing exciton is $\geq10\times$ larger, even larger than the diamagnetic shift coefficient of the 2s-Rydberg state.~\cite{Stier.2018} As we discuss in detail below, this observation points towards a much smaller reduced exciton mass of $X^{L}$ as compared to $X^0$, fully consistent with the quasi-linear exciton dispersion relation of $X^{L}$. 
Magnetic brightening experiments of dark excitons with \textit{in-plane} magnetic fields rely on Zeeman-induced coupling between bright and dark excitons, resulting in $B^2$-dependent brightening of the dark exciton.~\cite{Molas.2017} We observe the same $I^{X^{L}}(B)=\beta\cdot B^2$ brightening behavior for $X^{L}$, shown in Fig. \ref{figure:3}d, but with \textit{out-of-plane} magnetic fields. We attribute this brightening mechanism to $K^x$-induced coupling between massive and light-like excitons due to the kinetic energy acquired by the excitons through the diamagnetic term, $Q\propto B$. With increasing $Q$, excitons can populate higher-energy states in the dispersion relation, leading to a $\propto Q^2$ increase in the joint density of states for the light-like exciton - precisely matching our observed $\propto B^2$ increase of the $X^{L}$ emission.   
We now discuss how ab initio many-body theory confirms the interpretation of the emergent peak as $X^L$ based on the diamagnetic shift. 
Figure~\ref{fig:theory}a shows the exciton bandstructure computed using the GW and Bethe-Salpeter equation (GW-BSE) approach.\cite{Rohlfing.1998, Qiu.2015} The bright massive and light-like exciton branches are shown in solid black ($X^0$) and yellow ($X^L$), respectively. The splitting of the two bright excitonic bands originates from the non-local exchange interaction, as previously shown .~\cite{Qiu.2015, Qiu.2021} Dashed black lines denote dark excitons.\\ 
In general, the diamagnetic shift is a measure of exciton size divided by the reduced mass, both of which may depend on the exchange energy in particular in the presence of large magnetic fields. 
To investigate the $Q-$dependence of the diamagnetic shift, we calculate the exciton size of the two bright bands at the exciton momentum $Q=0.04\AA^{-1}$ (grey vertical line), the light-like exciton kinetic momentum at $B\approx30T$. Figures~\ref{fig:theory}d and ~\ref{fig:theory}e show the electron density distribution along an in-plane spatial coordinate (X-axis) for the linear- and parabolic-exciton branches, respectively. The hole is fixed on the tungsten atom at $X=0$. Shaded areas mark the range where 95\% of the electron density is localized, yielding the in-plane exciton diameter ($\approx 2\mathrm{r}_{\perp}$). We find that both excitons are of approximately the same spatial extent, suggesting that the large diamagnetic shift coefficient of $X^L$ is due to a very light reduced mass. We evaluate the $Q-$dependence of the coupling between the two branches and its effect on the diamagnetic shift. To do so, we calculate the single-particle diamagnetic shift ($\Sigma^{z}_{n\mathbf{k}}$)~\cite{Wozniak.2020}, using:

\begin{equation}\label{eq:dia_theory1}
    \hat{H}_{n\mathbf{k}} = -\frac{e^2\hbar^2B^2}{8m_0^3}\sum_m\frac{p^x_{nm\mathbf{k}}p^x_{mn\mathbf{k}}+p^y_{nm\mathbf{k}}p^y_{mn\mathbf{k}}}{\left(\epsilon_{n\mathbf{k}}-\epsilon_{m\mathbf{k}}\right)^2}=\Sigma^z_{n\mathbf{k}} B^2,
\end{equation}

where $n/m$ are band indices, $\mathbf{k}$ indexes electron/hole crystal momentum, $e$ is the elementary charge and $m_0$ the bare electron mass. $p^{x/y}$ are momentum matrix elements along the $x/y$ direction and $\epsilon$ is the electron/hole energy computed using a GW correction on top of density functional theory (DFT).~\cite{Amit2022}
From the single particle diamagnetic shift, we compute the many-body exciton diamagnetic shift coefficients using: 

\begin{equation}\label{eq:dia_theory2}
    \sigma^z_{S,\mathbf{Q}}=\left(\frac{m_0}{\mu}\right)^3\sum_{vc\mathbf{k}}{\left|A^\mathbf{Q}_{vc\mathbf{k}}\right|^{2}\left(\Sigma^z_{c\mathbf{k}}-\Sigma^z_{v\mathbf{k+Q}}\right)}.
\end{equation}

Here, $\mu$ is the effective exciton mass, $A^{\mathbf{Q}}_{vc\mathbf{k}}$ are the exciton spanning coefficients within the BSE basis set of occupied bands at the valence region, $v$, and unoccupied bands at the conduction region, $c$, as previously derived for many-body excitonic g-factors.~\cite{Amit2022,Deilmann2020PRL}
Figure~\ref{fig:theory}b shows the ratio of the many-body excitonic diamagnetic shift coefficients of the two bright exciton states as a function of $Q$, setting $\frac{\mu_L}{\mu_P}=1$.
This calculation shows that within our experimentally accessible magnetic field range, the diamagnetic shift ratio is only weakly affected by $Q$ and therefore our picture of a constant effective mass for the linear exciton branch is valid. Higher magnetic fields may alter this picture. 
 
In Fig.~\ref{fig:theory}c we calculate the ratio of the many-body exciton diamagnetic shift coefficients while manually tuning the ratio of $\frac{\mu_L}{\mu_p}$. Within our theory, we can explain the experimentally observed ratio of $\sigma_L/\sigma_P>20$ (see inset of Fig. \ref{fig:theory}c) to a $\geq 10\times$ lighter effective exciton mass of the linear branch as compared to the massive exciton ($\mu_P=0.2~m_e$\cite{Stier.2018}). \\
While the diamagnetic shift of $X^0$ is density-independent as expected for particles with  quadratic dispersion relations (Kohns theorem \cite{Kohn.1961}), $\sigma^z_{L}$ increases with increasing carrier density, leading to an increase of the ratio $\sigma^z_L/\sigma^z_P$, as shown in the inset in Fig.~\ref{fig:theory}c.
The linear exciton dispersion is predicted to originate from the $K^x$, a mechanism sensitive to excess electron density. While a complete understanding of this mechanism and its electron density dependence exceeds the scope of this manuscript, we argue that to leading order, Thomas-Fermi screening leads to a bandgap reduction with increased electron density.~\cite{Katoch2018} Within our experimental parameter space, this bandgap reduction is monotonic and therefore leads to an increase in $K^x\propto(E_c-E_v)^{-1}$, consistent with our experimentally observed increase of the rate of brightening, $\beta$, with density (see Fig. S3 in the SM). While a full determination of the exciton composition ($A^s_{vc\textbf{k}}$) as a function of reduced band gap is expected to be small but a theoretical challenge yet to be met, the observed density dependent increase in $\sigma^z_L/\sigma^z_P$ is consistent with an increased linear term in the $X^L$-dispersion relation, originating from an increased $K^x$.    

To conclude, we show high field magneto-spectroscopy of excitons in ML WSe$_2$, where we observe the emergence of a high energy peak from the neutral exciton, which we identify as the magnetically brightened light-like exciton, $X^L$. This peak features a valley Zeeman shift similar to $X^0$, but significantly smaller than $X^D$, consistent with a spin-singlet state. The observed very large diamagnetic shift of $X^L$ cannot be explained by Rydberg physics, but is consistent with the very light mass of the light-like exciton. The quadratic-in-$B$ brightening of this feature points towards a coupling mechanism that mixes $X^0$ and $X^L$ such that the latter gains oscillator strength consistent with a linear dispersion relation. The observed density dependence of the rate of brightening as well as the diamagnetic shift shows that only the light-like branch is affected by the excess electron density, as predicted by theory. Our results expand the plethora of excitons in monolayer semiconductors to particles obeying linear dispersion relations driven by electron-hole exchange interactions. 

A.D., P.J., J.J.F. and A.V.S. acknowledge the German Science Foundation (DFG) for financial support via the Priority Programmes SPP 2244 “2D Materials Physics of van der Waals heterobilayer”, as well as the clusters of excellence MCQST (EXS-2111) and e-conversion (EXS-2089). T. A. acknowledges support from the Azrieli Graduate Fellows Program. S. R.-A.  acknowledges support from a European Research Council (ERC) Starting Grant (No. 101041159). Computational resources were provided by the ChemFarm local cluster at the Weizmann Institute of Science. This work was partially supported by LNCMI, a member of the European Magnetic Field Laboratory (EMFL).

\bibliography{biblio}

\end{document}